\RequirePackage[hyphens]{url}
\documentclass[prd,twocolumn,a4paper,floatfix,nofootinbib,preprintnumbers,superscriptaddress]{revtex4-1}
\usepackage[utf8]{inputenc}
\usepackage{hyperref}
\usepackage{xcolor}
\usepackage{braket}
\usepackage[nolist,nohyperlinks]{acronym}
\usepackage{amsthm}
\usepackage{graphicx}
\usepackage{amsfonts}
\usepackage{booktabs}
\usepackage{siunitx}
\usepackage{mathtools}
\usepackage{multirow}
\usepackage[normalem]{ulem}
\usepackage{orcidlink}
\usepackage{natbib}
\usepackage{bm}
\usepackage{cleveref}
\usepackage{xcolor}
\usepackage{caption}
\usepackage{float}
\usepackage[nolist,nohyperlinks]{acronym}
\usepackage{tabularx}
\usepackage{physics}
\usepackage{subcaption}
\usepackage{placeins}
\usepackage[justification=RaggedRight]{caption}
\usepackage{transparent}
\usepackage[multiple]{footmisc}
\usepackage[dont-mess-around]{fnpct}
\usepackage{import}

\newcommand*\diff{\mathop{}\!\mathrm{d}}

\graphicspath{{./}}

\begin{document}

\title{Robust parameter estimation within minutes on gravitational wave signals from binary neutron star inspirals}

\author{Thibeau Wouters\thanks{t.r.i.wouters@uu.nl} \orcidlink{0009-0006-2797-3808}}
\email{t.r.i.wouters@uu.nl}
\affiliation{Institute for Gravitational and Subatomic Physics (GRASP), Utrecht University, Princetonplein 1, 3584 CC Utrecht, The Netherlands}
\affiliation{Nikhef, Science Park 105, 1098 XG Amsterdam, The Netherlands}
\author{Peter T. H. Pang \orcidlink{0000-0001-7041-3239}}
\affiliation{Institute for Gravitational and Subatomic Physics (GRASP), Utrecht University, Princetonplein 1, 3584 CC Utrecht, The Netherlands}
\affiliation{Nikhef, Science Park 105, 1098 XG Amsterdam, The Netherlands}
\author{Tim Dietrich \orcidlink{0000-0003-2374-307X}}
\affiliation{Institut f\"{u}r Physik und Astronomie, Universit\"{a}t Potsdam, Haus 28, Karl-Liebknecht-Str. 24/25, 14476, Potsdam, Germany}
\affiliation{Max Planck Institute for Gravitational Physics (Albert Einstein Institute), Am M\"{u}hlenberg 1, Potsdam 14476, Germany}
\author{Chris Van Den Broeck \orcidlink{0000-0001-6800-4006}}
\affiliation{Institute for Gravitational and Subatomic Physics (GRASP), Utrecht University, Princetonplein 1, 3584 CC Utrecht, The Netherlands}
\affiliation{Nikhef, Science Park 105, 1098 XG Amsterdam, The Netherlands}
\date{\today}

\begin{abstract}
\noindent The gravitational waves emitted by binary neutron star inspirals contain information on nuclear matter above saturation density. However, extracting this information and conducting parameter estimation remains a computationally challenging and expensive task. Wong et al.~introduced \textsc{Jim}~\cite{Wong:2023lgb}, a parameter estimation pipeline that combines relative binning and \textsc{jax} features such as hardware acceleration and automatic differentiation into a normalizing flow-enhanced sampler for gravitational waves from \ac{BBH} mergers. In this work, we extend the \textsc{Jim} framework to analyze gravitational wave signals from \ac{BNS} mergers with tidal effects included. We demonstrate that \textsc{Jim} can be used for full Bayesian parameter estimation of gravitational waves from \ac{BNS} mergers within a few tens of minutes, which includes the training of the normalizing flow and computing the reference parameters for relative binning. For instance, \textsc{Jim} can analyze GW170817 in $20$ minutes ($31$ minutes) of total wall time using the \texttt{TaylorF2} (\texttt{IMRPhenomD\_NRTidalv2}) waveform, and GW190425 in around $21$ minutes ($25$ minutes). We highlight the importance of such an efficient parameter estimation pipeline for several science cases as well as its ecologically friendly implementation of gravitational wave parameter estimation. 
\end{abstract}

\maketitle

\section{Introduction} \label{sec:intro}

\noindent \Acp{NS} are the remnants of core-collapse supernovae and consist of the densest matter ever observed in the Universe~\cite{Lattimer:2012nd,Ozel:2016oaf}. With densities up to a few times the nuclear saturation density $n_{\rm sat} = 2.7 \times 10^{14}{\rm g \ cm}^{-3}$, \acp{NS} are the perfect laboratories for studying the properties and behavior of ultra-dense matter. The \ac{EOS} relates the pressure, temperature, and energy density within the interior of \acp{NS} and is not completely understood~\cite{Ozel:2016oaf}. It is defined by the fundamental degrees of freedom within the \ac{NS} and the interactions among them. Each proposed \ac{EOS} uniquely dictates the global structure of \acp{NS}, influencing their masses, radii, and tidal deformabilities~\cite{Hinderer:2009ca,Damour:2009vw,Lindblom1992-zb}. Thus, astronomical observations of \acp{NS} allow one to constrain the \ac{EOS} in \acp{NS}~\cite{Flanagan:2007ix, DelPozzo:2013ala,Agathos:2015uaa, Demorest:2010bx, Antoniadis:2013pzd, Arzoumanian:2017puf, Cromartie:2019kug, Miller:2019cac, Riley:2019yda, Miller:2021qha, Riley:2021pdl}.

Since the first detection of \acp{GW} from a \ac{BNS} merger in 2017~\cite{LIGOScientific:2017vwq}, by Advanced LIGO~\cite{LIGOScientific:2014pky} and Advanced Virgo~\cite{VIRGO:2014yos}, \ac{GW} astronomy has become an important channel for astronomical observation of \acp{NS}. An essential step for extracting physics from the \ac{GW} data is performing \ac{PE} on it.

Within the community, multiple \ac{CPU}-based \ac{PE} software packages have been developed, which include \textsc{LALInference}~\cite{Veitch:2014wba}, \textsc{PyCBC Inference}~\cite{Biwer:2018osg} and \textsc{Bilby}~\cite{Ashton:2018jfp,Romero-Shaw:2020owr,Smith:2019ucc}. These packages are robust and have been used for analyzing multiple \ac{GW} events \cite{LIGOScientific:2017vwq,LIGOScientific:2020aai}. Yet, they are also known for being computationally expensive and having a large carbon footprint, especially for analyzing \ac{BNS} events. For these reasons, a lot of effort has been put into developing methods to speed up \ac{PE}, e.g., relative binning~\cite{Cornish:2010kf,Zackay:2018qdy} and \ac{ROQ}~\cite{Canizares:2013ywa, Canizares:2014fya}, which show substantial improvements~\cite{Krishna:2023bug,Morisaki:2023kuq}.

In addition, there have been proposals to accelerate \ac{PE} through the use of \ac{GPU}-based software or \ac{ML} techniques. One such technique, known as normalizing flows, has been gaining popularity in the field. For instance, \textsc{Dingo} uses normalizing flows pretrained on simulated data to approximate the posterior distribution~\cite{Dax:2021tsq, Dax:2022pxd,Bhardwaj:2023xph}. Similarly, normalizing flows have been used to accelerate nested sampling for \ac{GW} data analysis by \textsc{nessai}~\cite{Williams:2021qyt, Williams:2023ppp}. Recently, a \ac{GPU}-based \ac{PE} software, \textsc{Jim}~\cite{Wong:2023lgb}, has been introduced, which accelerates \ac{PE} with \ac{MCMC} samplers using normalizing flows. In Ref.~\cite{Wong:2023lgb}, the authors demonstrated that \textsc{Jim} can conduct \ac{PE} on \ac{GW} signals within minutes of total wall time on a single \ac{GPU}. 

In this paper, we extend the capabilities of \textsc{Jim} to analyze \acp{GW} from \ac{BNS} mergers, in particular to also infer the tidal deformabilities of \ac{GW} signals. The updated \textsc{Jim} framework is crucial for low-latency \ac{PE}, necessary for efficient telescopes' responses on \ac{BNS} detections, for handling the \ac{PE} for a large number of \ac{BNS} events, and to aid the multi-messenger analysis on \ac{BNS} via integration with a nuclear physics multi-messenger framework, e.g., \textsc{NMMA}~\cite{Pang:2022rzc}. As we show in more detail below, \textsc{Jim} performs these efficient runs at a lower carbon footprint compared to other existing pipelines.

This paper is structured as follows. In Sec.~\ref{sec:methods}, we give an overview of \textsc{Jim} and our methods, with their validation shown in Sec.~\ref{sec:validation}. In Sec.~\ref{sec:result}, we apply our methods on the two \ac{BNS} \ac{GW} signals detected to date, namely GW170817~\cite{LIGOScientific:2017vwq} and GW190425~\cite{LIGOScientific:2020aai}. We compare our work to other methods and discuss implications for future work in Sec.~\ref{sec:discussion}. Concluding remarks and future perspectives are provided in  Sec.~\ref{sec:conclusions}.

\section{Methods} \label{sec:methods}
\subsection{Parameter estimation}
\noindent Based on Bayes' theorem, the posterior distribution of the source parameters $\boldsymbol{\theta}$ of a \ac{GW} signal $d$, denoted by $p(\boldsymbol{\theta}|d)$, is given by 
\cite{Veitch:2009hd}
\begin{equation}
\begin{aligned}
    p(\boldsymbol{\theta}|d) = \frac{p(d|\boldsymbol{\theta})p(\boldsymbol{\theta})}{p(d)} \,,
\end{aligned}
\end{equation}
where $p(d|\boldsymbol{\theta})$ is the likelihood function, $p(\boldsymbol{\theta})$ is the prior probability distribution and $p(d)$ is the Bayesian evidence. Under the assumption of stationary Gaussian noise, the log-likelihood function for the waveform $h(\boldsymbol{\theta})$ is given by
\begin{equation}\label{eq: likelihood function}
    \log p(d|\boldsymbol{\theta}) = - \frac12 \left\langle d - h(\boldsymbol{\theta}), d - h(\boldsymbol{\theta}) \right\rangle + {\rm constant},
\end{equation}
where the inner product $\langle a, b \rangle$ is defined as
\begin{equation}\label{eq:noise_weighted_inner_product}
    \langle a, b \rangle = 4 \real \int_{f_{\rm low}}^{f_{\rm high}} \diff f \frac{\tilde{a}(f) \tilde{b}^*(f)}{S_n(f)} \, , 
\end{equation}
with $S_n(f)$ being the one-sided \ac{PSD}, $\tilde{x}(f)$ being the Fourier transform of $x(t)$, and the asterisk denoting the complex conjugate. In this work, the frequency range of the integral is chosen such that it covers the frequency span of the signal, which for \ac{BNS} signals is usually taken as $[20, 2048]$ Hz. A complete list of parameters and corresponding definitions can be found in Ref.~\cite{Romero-Shaw:2020owr}. The notation used in this work can be found in Tab.~\ref{tab:parameter_priors}. 

\subsection{Waveform approximants}

\noindent Several families of waveform approximants have been developed and are used in \ac{GW} data analysis. These families are mainly divided into the \ac{PN} waveforms~\cite{Buonanno:2009zt, Blanchet:2013haa}, the effective one-body waveforms~\cite{Buonanno:1998gg, Buonanno:2000ef, Damour:2009zoi}, such as \texttt{TEOBResumS}~\cite{Bernuzzi:2014owa,Nagar:2018zoe,Akcay:2018yyh} or \texttt{SEOBNRv4T}~\cite{Hinderer:2016eia,Steinhoff:2016rfi} for \ac{BNS} inspirals, the \ac{IMRPhenom} waveforms, such as \texttt{IMRPhenomD} for \ac{BBH} inspirals~\cite{Husa:2015iqa, Khan:2015jqa}, or surrogate models based on numerical-relativity simulations~\cite{Varma:2019csw}. In this work, we will limit ourselves to the frequency domain approximants \texttt{TaylorF2}, which is a \ac{PN} approximant, and \texttt{IMRPhenomD\_NRTidalv2} \cite{Dietrich:2017aum, Dietrich:2019kaq}, which are both frequently used in the analysis of \ac{BNS} events.  We do not consider the recently introduced  \texttt{IMRPhenomXAS\_NRTidalv2}~\cite{Colleoni:2023czp} or \texttt{NRTidalv3} models~\cite{Abac:2023ujg}, since these waveform models are not released at the time of writing. 

\subsection{Relative binning}\label{sec: relative binning}

\noindent Relative binning~\cite{Zackay:2018qdy, Cornish:2021lje, Krishna:2023bug} is a technique to speed up the evaluation of the likelihood function given in Eq.~\eqref{eq: likelihood function}. Given a reference parameter $\boldsymbol{\theta}_{\rm ref}$, the ratio $r(f)$ of the waveform of an arbitrary parameter $\boldsymbol{\theta}$ against the reference waveform is given by
\begin{equation}\label{eq: relative binning ratio}
\begin{aligned}
     r(f) &= \frac{h(f;\boldsymbol{\theta})}{h(f;\boldsymbol{\theta}_{\rm ref})}\\
     &= \left| \frac{A(f)}{A_{\rm ref}(f)} \right| e^{-i (\Psi(f) - \Psi_{\rm ref}(f))} \, .
\end{aligned}
\end{equation}
The above ratio can be approximated by piecewise linear functions, i.e.,
\begin{equation}
    r(f) \approx
    \begin{cases}
    r_0(b_{1}) + r_1(b_1)(f - f_{\rm m}(b_1)), & f \in b_1\\
    r_0(b_{2}) + r_1(b_2)(f - f_{\rm m}(b_2)), & f \in b_2\\
    \qquad \qquad \qquad \vdots & \\
    r_0(b_{n}) + r_1(b_n)(f - f_{\rm m}(b_n)). & f \in b_n\\
    \end{cases}
\end{equation}
Thus, the ratio is approximated with frequencies split into bins $\{b_i\}$, with central frequencies $\{f_{\rm m}(b_i)\}$. Such an approximation can reach an arbitrary accuracy provided that a sufficiently high number of bins $b_i$ is used. However, to speed up the \ac{PE}, one would aim for approximating the above ratio with the least number of bins. The relative binning method assumes that the region of parameter space with a non-negligible likelihood overlaps with the region for which the waveforms only differ from the best-fit waveform by small perturbations. Thus the ratio $r(f)$ is a smooth function that can be well approximated with a low number of bins.

The bin placement often follows the approach described in Ref.~\cite{Zackay:2018qdy}. In this scheme, one assumes that the variations in the ratio of amplitudes can be neglected. The error on the ratio is therefore determined by the deviations in the phase, and one can place the bin edges in such a way to ensure that the deviation in the phase across a bin is below a certain desired threshold. This deviation is approximated with an ansatz of the phase based on \ac{PN} theory to remove the dependence on the waveform's parameters. We refer readers to Refs.~\cite{Zackay:2018qdy, Krishna:2023bug} for further details.

\begin{figure*}[htpb]
    \centering
    \tiny
    \def\svgwidth{\columnwidth}
    \resizebox{\textwidth}{!}{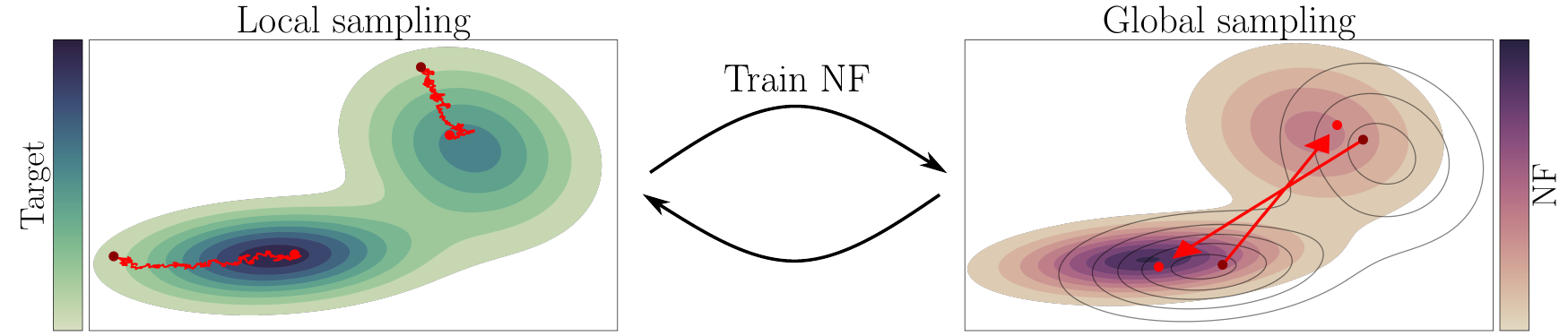}

    \normalsize
    \caption{Schematic overview of the training loop of the \textsc{flowMC} sampler. Each loop starts with running the local sampler. For the local sampling, we use the \ac{MALA} algorithm, which exploits the gradient of the target distribution (green and gray lines) to evolve the Markov chains (red). With the samples obtained from the local sampler, a normalizing flow (NF) is trained to approximate the distribution of the Markov chains. During the global sampling phase, we use the density learned by the NF (purple) as a proposal. We accept or reject the proposed samples (red) with a Metropolis-Hastings step, which relies on both the proposal density as well as the target density. This local-global procedure is repeated until the NF has converged. Afterward, we perform a fixed number of production loops, where the weights of the NF are frozen and the local and global sampler output the final production samples.}
    \label{fig: flowMC diagram}
\end{figure*}

\subsubsection{Caveats for applying relative binning with \texttt{NRTidal}-family waveforms}\label{sec: relative binning nrtidal caveat}
\vspace{\parskip}

The above assumption regarding the amplitude ratio holds for a majority of waveform models (e.g., \texttt{TaylorF2}). However, it is invalid for the \texttt{IMRPhenomD\_NRTidalv2} waveform model. This waveform model modifies its underlying point-particle baseline, \texttt{IMRPhenomD}, with closed-form expressions for the tidal contributions to the phase and amplitude. These are modeled by Padé approximants, such that the \ac{PN} ansatz for the phase used in placing the bin edges is no longer a valid approximation. Moreover, \texttt{IMRPhenomD\_NRTidalv2} additionally applies a Planck taper~\cite{McKechan:2010kp} to the waveform so that its amplitude transitions smoothly to zero in the range ${[f_{\rm merger}, 1.2f_{\rm merger}]}$, where the merger frequency $f_{\rm merger}$ depends on the intrinsic source parameters and was estimated through a phenomenological relation~\cite{Dietrich:2019kaq}. For larger deviations from the reference parameters, this causes the ratio of amplitudes in Eq.~\eqref{eq: relative binning ratio} to vary significantly, causing a breakdown of the relative binning scheme. Moreover, we observed that, for some parameter regimes for which the reference waveform has a low merger frequency, the ratio of amplitudes becomes numerically unstable. Therefore, \textsc{Jim} generates the reference waveform without the tapering window when analyzing signals with the \texttt{IMRPhenomD\_NRTidalv2} waveform model, whereas we still include the taper in the sampled waveforms. The injected signals of our injections in Sec.~\ref{sec:validation} also apply the taper. We note that \textsc{bilby}, contrary to \textsc{Jim}, includes the taper in the reference waveform.

\subsubsection{Reference parameters}
Relative binning requires a set of reference parameters as input that, ideally, lie close to the best-fit parameters. When analyzing simulated signals, we set the reference parameters equal to the injected parameters. For analyzing real events, we first perform a heuristic search to obtain an estimate of the maximum likelihood parameters. More specifically, we employ the covariance matrix adaptation evolution strategy (CMA-ES)~\cite{CMA_ES_reference_paper} as implemented in \textsc{evosax}~\cite{lange2022evosax} before running \textsc{Jim} to obtain the reference parameters used in the \ac{PE} run. When searching for a set of reference parameters with the \texttt{IMRPhenomD\_NRTidalv2} waveform, we include its amplitude taper.

\subsection{\textsc{Jim}}

\textsc{Jim}~\cite{Wong:2023lgb} is a \ac{PE} pipeline to analyze \ac{GW} events involving \ac{BBH} mergers that is able to run on \ac{CPU}s, \ac{GPU}s or tensor processing units (TPUs). In Ref.~\cite{Wong:2023lgb}, the authors demonstrated \textsc{Jim}'s capability of performing \ac{PE} in the order of minutes of total wall-time, including training of the normalizing flow. In this work, we extend \textsc{Jim}'s capability of performing \ac{PE} on \ac{GW} signals from \ac{BNS} inspirals with tidal effects included.

\textsc{Jim} is implemented in \textsc{jax}~\cite{frostig2018compiling}, a high-performance numerical computation library. \textsc{jax} has a number of desirable features for \ac{GW} data analysis, such as i) automatic differentiation, allowing the use of, e.g., gradient-based \ac{MCMC} samplers, ii) native support for hardware accelerators, such as \acp{GPU} or TPUs, and iii) just-in-time (JIT) compilation to further accelerate the execution of the code.

For our waveform generator, we make use of \textsc{ripple}~\cite{Edwards:2023sak}, a \textsc{jax} package for differentiable waveform approximants. To analyze \ac{BNS} signals, we extend \textsc{ripple} with the \texttt{TaylorF2} and \texttt{IMRPhenomD\_NRTidalv2} waveform models.\footnote{While \texttt{TaylorF2} and \texttt{IMRPhenomD\_NRTidalv2} already have an existing \textsc{jax} implementation in \textsc{gwfast}~\cite{Iacovelli:2022mbg}, we choose to provide an independent implementation in \textsc{ripple} in order to interface more easily with \textsc{Jim} and \textsc{flowMC}.}\footnote{We note that Ref.~\cite{Chia:2023tle} has used \texttt{TaylorF2} with \textsc{Jim} for \ac{PE}. However, here we provide an extensive validation of the implementation and provide detailed benchmarks.}

For the \ac{MCMC} sampler, we employ \textsc{flowMC}, a \textsc{jax}-based \ac{MCMC} sampler that makes use of gradient-based samplers and is enhanced by normalizing flows~\cite{Wong:2022xvh, Gabrie:2021tlu}. The \textsc{flowMC} sampler is summarized by the diagram in Fig.~\ref{fig: flowMC diagram}. \textsc{flowMC} combines a local and global proposal distribution to improve the sampler's efficiency. For the local sampler, we use the Metropolis-adjusted Langevin algorithm (MALA) routine \cite{grenander1994representations}, which exploits the gradient of the posterior distribution to evolve the Markov chains. The global sampler is parameterized by a \ac{NF}, which is implemented with \textsc{equinox}~\cite{kidger2021equinox}. \acp{NF} are deep generative models that offer tractable approximations of complex probability distributions, allowing for efficient and precise sampling as well as density evaluation. In \textsc{flowMC}, the \ac{NF} is trained from the samples generated by the local sampler and subsequently used as proposal distribution for the global sampling phase.

The sampler settings can be either optimized for speed or accuracy. In this work, we use settings that focus on the latter. For instance, to improve the robustness of the sampler, we use a stopping threshold. Once the \ac{NF} achieves a preset threshold value for the mean acceptance rate during the Metropolis-Hastings steps, the \ac{NF} is frozen, and the final production samples are produced by only running the local and global samplers (known as the production loop) for a fixed amount of epochs. Training of the \ac{NF} dominates the wall time, while the production loop only takes $1$-$2$ minutes. 

\section{Validation} \label{sec:validation}

In this section, we validate our methods before applying them to real \ac{GW} data in Sec.~\ref{sec:result}. In particular, we check the accuracy of the waveforms implemented in \textsc{ripple} and verify the robustness of our pipeline via an injection-recovery test.

\subsection{Accuracy of {\normalfont \textsc{ripple}} waveforms}
We verify the correctness of our waveform generators by comparing them against the \textsc{LALsuite} \cite{lalsuite} implementation. This is done by computing the mismatch $M$ between the waveforms generated by the two implementations, which is defined by
\begin{equation}
    M(h_1, h_2) \equiv 1 - \max_{\Delta t_c, \Delta \phi_c} \frac{\langle h_1, h_2 \rangle}{\sqrt{\langle h_1, h_1 \rangle \langle h_2, h_2 \rangle}} \, ,
\end{equation}
where $\Delta t_c$ and $\Delta \phi_c$ represent the difference between the time of coalescence and the phase of coalescence of the two waveforms, respectively.

\begin{figure}
    \centering
    \vspace{1mm}
\includegraphics[width=\linewidth]{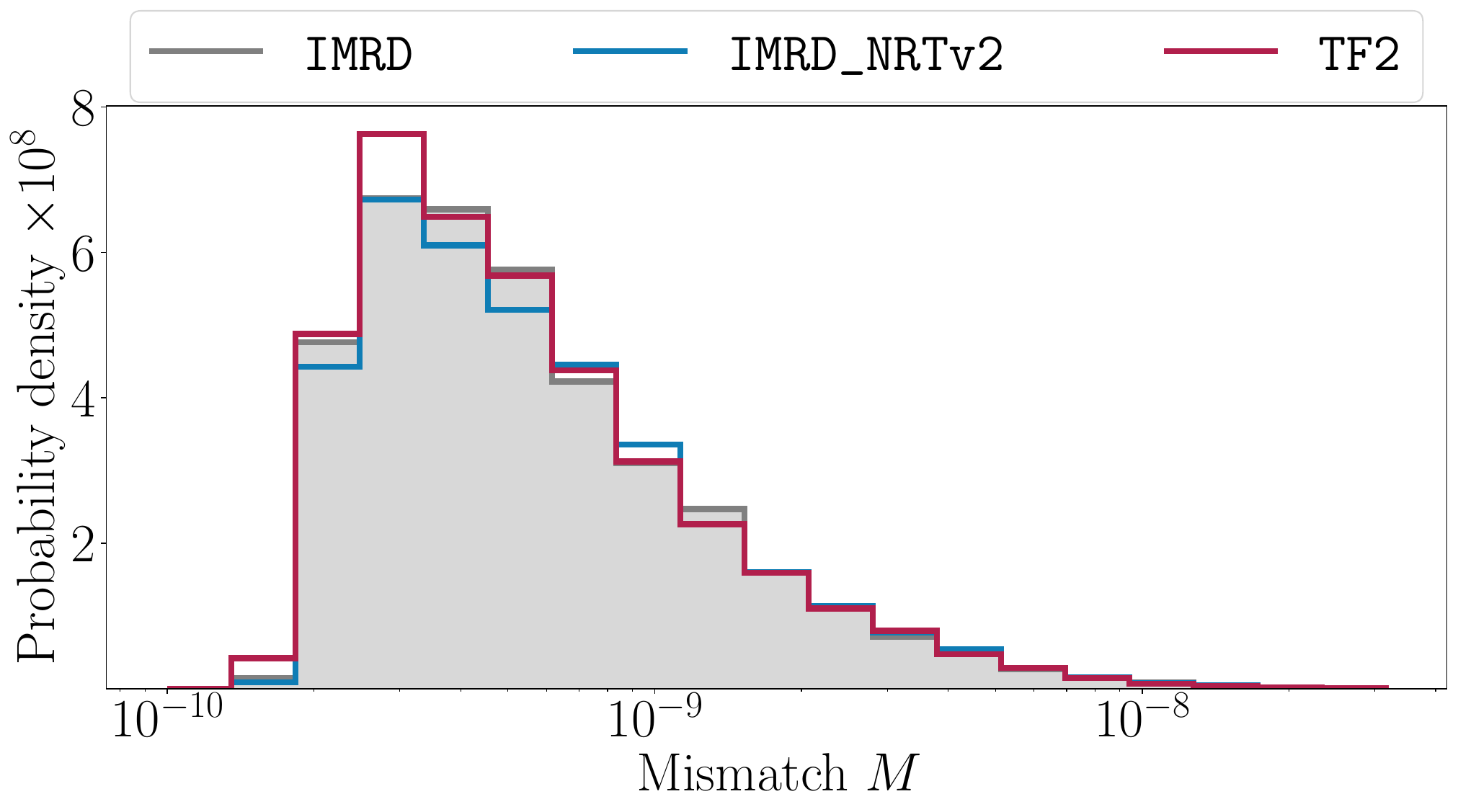}
    \caption{Distribution of mismatch values between \textsc{ripple} and \textsc{LALsuite} implementations of the \texttt{IMRPhenomD} (\texttt{IMRD}), \texttt{IMRPhenomD\_NRTidalv2} (\texttt{IMRD\_NRTv2}) and \texttt{TaylorF2} (\texttt{TF2}) waveforms. For both newly added waveform models, the probability mass peaks at a mismatch below $10^{-8}$, indicating consistency with the \textsc{LALsuite} implementation.}
    \label{fig:ripple_mismatch}
\end{figure}

In Fig.~\ref{fig:ripple_mismatch}, the distribution of the mismatch between the \textsc{ripple} waveforms against the corresponding \textsc{LALsuite} implementation is shown. Specifically, we have sampled ${10000}$ parameters from the distributions shown in Tab.~\ref{tab:mismatch_parameters}. The mismatch is computed on a frequency grid in the range $[20, 2048]$ Hz with spacing $\Delta f = 1/T$, where we take the duration $T$ to be $128$ s, and with the design sensitivity of Advanced LIGO~\cite{LIGOScientific:2014pky} used for the \ac{PSD}. The mismatches for the \texttt{IMRPhenomD\_NRTidalv2} waveform are computed with the Planck taper still included in the \textsc{ripple} implementation. For comparison, we also show the mismatches for \textsc{ripple}'s \ac{BBH} waveform \texttt{IMRPhenomD} for the same parameter ranges, excluding the tidal deformabilities. All \textsc{ripple} waveform samples have mismatches $M \lesssim 10^{-8}$ against their \textsc{LALsuite} counterpart, making them indistinguishable for parameter estimation~\cite{Purrer:2019jcp}. 

\begin{table}
    \centering
    \renewcommand{\arraystretch}{1.0}
    \begin{tabular*}{0.925\linewidth}{@{\extracolsep{\fill}} l l}
        \hline\hline
        Parameter & Range \\
        \hline
        Component masses & $[0.5M_{\odot}, 3M_{\odot}]$\\
        Component aligned spins & $[-0.05, 0.05]$\\
        Dimensionless tidal deformabilities &  $[0, 5000]$\\
        Inclination angle & $[0, \pi]$\\
        \hline\hline
    \end{tabular*}
    \caption{Parameter ranges for the mismatch calculation. All parameters are distributed uniformly in the specified ranges.}
    \label{tab:mismatch_parameters}
\end{table}

\subsection{Injection-recovery test}

To demonstrate the robustness of our \ac{PE} pipeline, we perform injection-recovery tests and report the results in a \ac{p-p} plot~\cite{cook2006validation}. We inject simulated \ac{GW} signals into realizations of noise from a detector network and conduct \ac{PE} on the simulated data with \textsc{Jim}. After performing a batch of such analyses, we calculated the credible level at which each true parameter appears in its marginal posterior distribution. If the pipeline delivers unbiased estimates of the source properties, it is expected that the true parameters occur in the $x\%$ credible interval $x\%$ of the time. Therefore, the cumulative distribution of the credible level should trend along the diagonal, which is often checked visually with a \ac{p-p} plot.

\begin{table*}
    \centering
    \renewcommand{\arraystretch}{1.25}
    \begin{tabular*}{0.8\linewidth}{@{\extracolsep{\fill}} l l l l l}
    \hline \hline
    Parameter & Description & Injection & GW170817 & GW190425 \\ \hline
    $\mathcal{M}$ & detector-frame chirp mass $[M_{\odot}]$ & $[0.88, 2.61]$ & $[1.18, 1.21]$ & $[1.485, 1.490]$ \\
    $q$ & mass ratio & $[0.5, 1]$ & $[0.125, 1]$ & $[0.125, 1]$ \\ 
    $\chi_1, \chi_2$ & aligned spins & $[-0.05, 0.05]$ & $[-0.05, 0.05]$ & $[-0.05, 0.05]$ \\
    $\Lambda_1, \Lambda_2$ & dimensionless tidal deformabilities & $[0, 5000]$ & $[0, 5000]$ & $[0, 5000]$ \\
    $d_L$ & luminosity distance $[\rm{Mpc}]$ & $[30, 300]$ & $[1, 75]$ & $[1, 500]$ \\
    $t_c$ & coalescence time $[\rm{s}]$ & $[-0.1, 0.1]$ & $[-0.1, 0.1]$ & $[-0.1, 0.1]$ \\
    $\phi_c$ & coalescence phase & $[0, 2\pi]$ & $[0, 2\pi]$ & $[0, 2\pi]$ \\
    $\cos \iota$ & cosine of inclination angle & $[-1, 1]$ & $[-1, 1]$ & $[-1, 1]$ \\
    $\psi$ & polarization angle & $[0, \pi]$ & $[0, \pi]$ & $[0, \pi]$ \\
    $\alpha$ & right ascension & $[0, 2\pi]$ & $[0, 2\pi]$ & $[0, 2\pi]$ \\
    $\sin \delta$ & sine of declination & $[-1, 1]$ & $[-1, 1]$ & $[-1, 1]$ \\
    \hline \hline
    \end{tabular*}
    \caption{Parameters used and their corresponding prior ranges for our analyses. All priors considered in this work are uniform priors with the specified range.}
    \label{tab:parameter_priors}
\end{table*}

\begin{figure*}
     \includegraphics[width=0.49\textwidth]{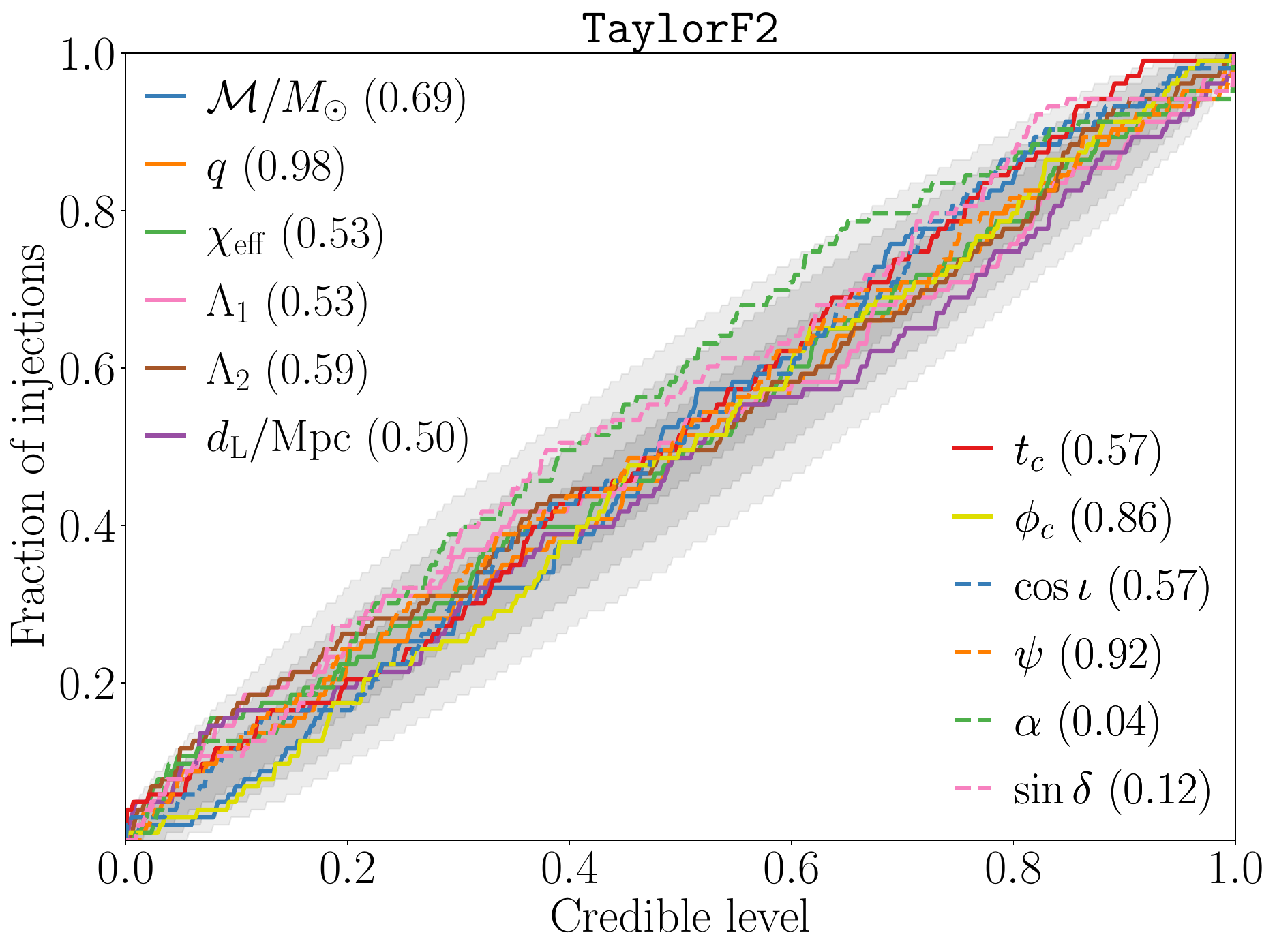}
     \includegraphics[width=0.49\textwidth]{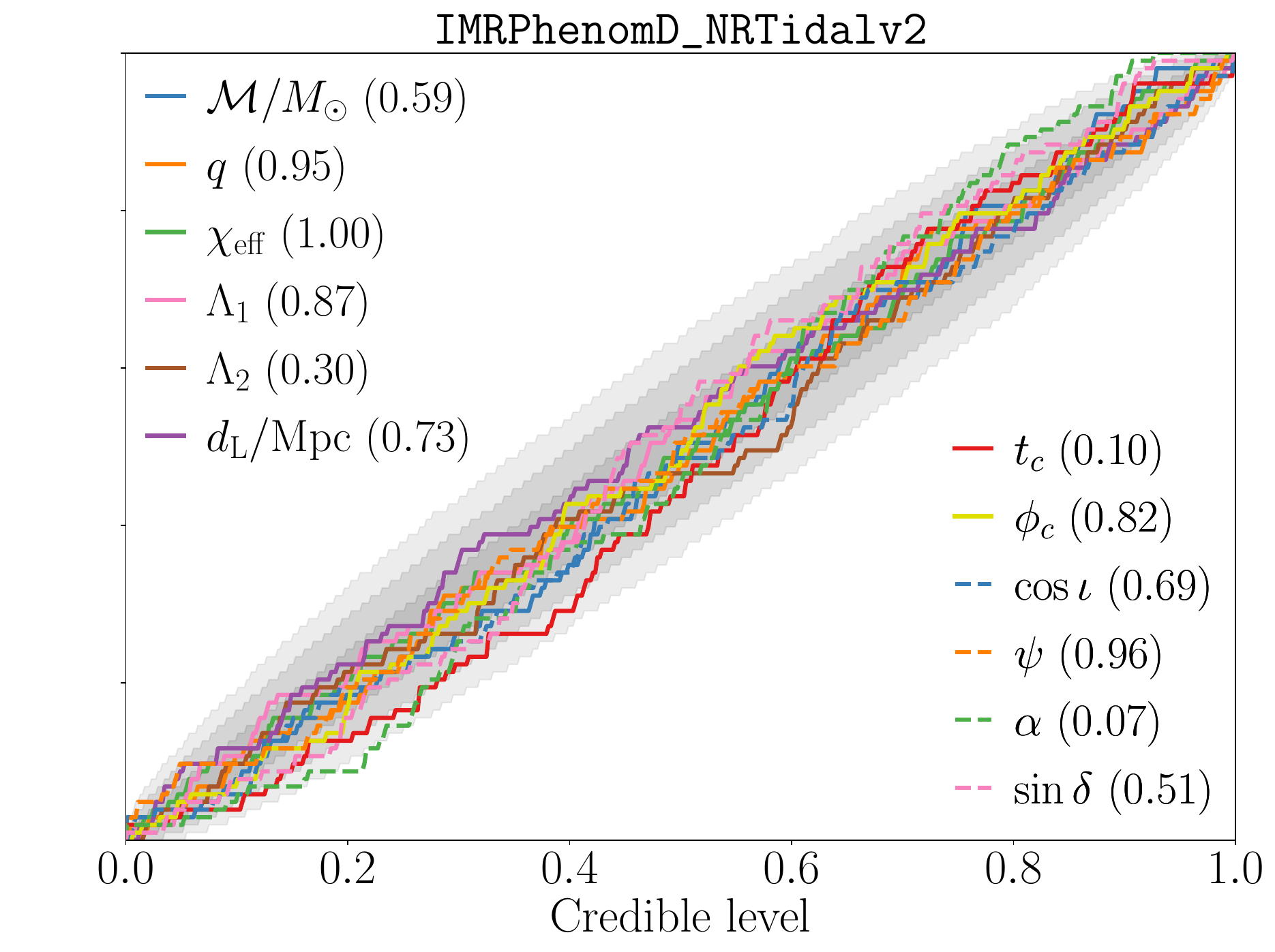}
    \caption{\ac{p-p} plot for the injections done with \texttt{TaylorF2} (left) and \texttt{IMRPhenomD\_NRTidalv2} (right), each created from 100 injections. For both of the waveform models, the \ac{p-p} plots agree well with the diagonal, demonstrating the robustness of the \textsc{Jim} pipeline on analyzing \ac{BNS} signals. The shaded regions indicate the $1\sigma$, $2\sigma$, and $3\sigma$ credible levels due to finite sample size.}
    \label{fig:pp plots}
\end{figure*}

For both waveforms considered here, we create 100 \ac{GW} events by sampling source parameters uniformly over the ranges given in Tab.~\ref{tab:parameter_priors}. We reject sampled injection parameters that result in a signal-to-noise ratio below $12$.\footnote{In principle, this introduces a small bias due to a difference between the prior distributions used for injection and recovery. Since we have found this to be a minor effect, we neglect this bias in this work.} The duration is set to $128$ s and the frequency range to $[20, 2048]$ Hz. We use $100$ bins for relative binning. The synthetic \ac{GW} signal is injected into a network consisting of the Advanced LIGO  \cite{LIGOScientific:2014pky} and Advanced Virgo \cite{VIRGO:2014yos} detectors at their design sensitivities. We evolve ${1000}$ Markov chains and stop the \textsc{flowMC} training phase once the \ac{NF} achieves an average acceptance rate of $20\%$. Instead of the individual aligned spins, we show the effective spin $\chi_{\rm eff}$, defined by
\begin{equation}\label{eq:effective spin}
    \chi_{\rm eff} = \frac{m_1 \chi_1 + m_2 \chi_2}{m_1 + m_2} \, .
\end{equation} The resulting \ac{p-p} plots are shown in Fig.~\ref{fig:pp plots}. The combined $p$-values are $0.74$ and $0.85$, for the injections made with the \texttt{TaylorF2} and \texttt{IMRPhenomD\_NRTidalv2} waveforms, respectively. Since the \ac{p-p} plots also trend along the diagonal, we conclude that our pipeline is robust.

\section{Analysis on GW170817 and GW190425} \label{sec:result}
With the robustness of our pipeline demonstrated, we apply \textsc{Jim} to the two \ac{BNS} events observed so far: GW170817~\cite{LIGOScientific:2017vwq} and GW190425~\cite{LIGOScientific:2020aai}\footnote{While GW170817 has been confidently verified to be a \ac{BNS} merger, there have been proposals that GW190425 could be a neutron-star–black-hole merger~\cite{Foley:2020kus, Han:2020qmn,Kyutoku:2020xka,Hinderer:2018pei, Coughlin:2019kqf}.}. For both events, we take the publicly available data from GWOSC~\cite{LIGOScientific:2019lzm, KAGRA:2023pio} as input. The priors used in the analysis are reported in Tab.~\ref{tab:parameter_priors}.

For comparison, we have conducted the \ac{PE} with both \texttt{TaylorF2} and \texttt{IMRPhenomD\_NRTidalv2} using both \textsc{Jim} and \textsc{Parallel-Bilby} (\textsc{pBilby})~\cite{Smith:2019ucc} to verify the accuracy of \textsc{Jim}. As mentioned in Sec.~\ref{sec: relative binning}, \textsc{Jim} utilizes relative binning, employing \textsc{evosax} to search for the reference parameters. We evolve ${1000}$ Markov chains and stop the \textsc{flowMC} training phase once the \ac{NF} achieves an average acceptance rate of $10\%$. We use $200$ bins ($1000$ bins), when using the \texttt{TaylorF2} (\texttt{IMRPhenomD\_NRTidalv2}) waveform, respectively. The resulting posterior samples are compared in Sec.~\ref{sec:posterior_comparison}, and the run times are compared in Sec.~\ref{sec:wall time}.

\subsection{Posterior comparison}
\label{sec:posterior_comparison}

In~\Cref{fig:GW170817_TaylorF2,fig:GW170817_NRTidalv2,fig:GW190425_TaylorF2,fig:GW190425_NRTidalv2} in Appendix B, we show the visual comparison between the posterior samples obtained from \textsc{Jim} and \textsc{pBilby} and find agreement between the posteriors. To make a quantitative statement, we report the \ac{JS} divergences \cite{lin1991divergence} in bits between the marginal distributions in Tab.~\ref{tab: JS divergences events}. The \ac{JS} divergences lie between $0$ and $1$ and the highest \ac{JS} divergence is $0.026363$ bits ($0.010634$ bits) for GW170817 (GW190425). To get a sense of these numbers, we can compare them against \ac{JS} divergences of two numerical experiments. First, in the final column of Tab.~\ref{tab: JS divergences events}, we show the maximal \ac{JS} divergence for each parameter found after varying the seed of the \ac{RNG} for the analysis of GW190425 with \texttt{TaylorF2} $10$ times. This provides a lower bound on the expected \ac{JS} divergence when comparing posterior distributions, as we expect that fluctuations in the \ac{RNG} will only marginally affect the posteriors. Second, Ref.~\cite{Romero-Shaw:2020owr} proposed the criterion to consider two empirical distributions to be identical if their \ac{JS} divergence is below $0.0022$ bits. While some of the \ac{JS} divergences in Tab.~\ref{tab: JS divergences events} are above this threshold, we can conclude that there is agreement between the results of the \textsc{Jim} and \textsc{pBilby} pipelines.

\begin{table}[h]
    \centering
    \def\arraystretch{1.2}
    \begin{tabular*}{0.95\linewidth}{@{\extracolsep{\fill}} l  c  c  c  c | c }
\hline\hline
& \multicolumn{2}{ c }{GW170817} & \multicolumn{2}{ c }{GW190425} & \\

 & \texttt{TF2} & \texttt{NRTv2} & \texttt{TF2} & \texttt{NRTv2} & RNG \\

$\mathcal{M}$ & $0.001725$ & $0.001553$ & $0.003557$ & $0.001626$ & $0.000261$ \\
$q$ & $0.005212$ & $0.003418$ & $0.004837$ & $0.003102$ & $0.000242$ \\
$\chi_1$ & $0.005633$ & $0.003815$ & $0.002794$ & $0.001512$ & $0.000491$ \\
$\chi_2$ & $0.003030$ & $0.003373$ & $0.002416$ & $0.001533$ & $0.000247$ \\
$\Lambda_1$ & $0.001062$ & $0.002802$ & $0.008556$ & $0.000690$ & $0.000363$ \\
$\Lambda_2$ & $0.000559$ & $0.003725$ & $0.005808$ & $0.001395$ & $0.000477$ \\
$d_L$ & $0.001544$ & $\bm{0.026363}$ & $0.001273$ & $0.008755$ & $0.000969$ \\
$\phi_c$ & $0.003500$ & $0.013961$ & $0.003338$ & $0.007819$ & $0.000281$ \\
$\cos \iota$ & $0.001615$ & $0.022125$ & $0.006400$ & $0.009580$ & $0.000289$ \\
$\psi$ & $0.004048$ & $0.015805$ & $0.001516$ & $0.005477$ & $0.000203$ \\
$\alpha$ & $\bm{0.014008}$ & $0.001896$ & $\bm{0.009822}$ & $\bm{0.010634}$ & $0.001624$ \\
$\sin \delta$ & $0.009570$ & $0.000735$ & $0.008934$ & $0.008896$ & $0.000640$ \\
\hline\hline
    \end{tabular*}
    \caption{Jensen-Shannon divergences (in bits) between the marginal posterior obtained for GW170817 and GW190425 using \texttt{TaylorF2} and \texttt{IMRPhenomD\_NRTidalv2} with \textsc{Jim} and \textsc{pBilby}, with the highest value of each comparison in bold. The \ac{JS} divergences are bound between $[0, 1]$. The final column shows the maximal deviation one obtains from fluctuations in the random number generator (RNG).}
    \label{tab: JS divergences events}
\end{table}

%%% OLD TABLE HERE, TODO: remove it!
% \begin{table}[h]
%     \centering
%     \def\arraystretch{1.2}
%     \begin{tabular*}{0.9\linewidth}{@{\extracolsep{\fill}} l  c  c  c  c }
% \hline\hline
% & \multicolumn{2}{ c }{GW170817} & \multicolumn{2}{ c }{GW190425} \\
%  & \texttt{TF2} & \texttt{NRTv2} & \texttt{TF2} & \texttt{NRTv2} \\ \hline

% $\mathcal{M}$ & $0.001725$ & $0.000516$ & $0.003557$ & $0.001884$ \\
% $q$ & $0.005212$ & $0.007894$ & $0.004837$ & $0.003883$ \\
% $\chi_1$ & $0.005633$ & $0.004301$ & $0.002794$ & $0.001904$ \\
% $\chi_2$ & $0.003030$ & $0.002671$ & $0.002416$ & $0.001400$ \\
% $\Lambda_1$ & $0.001062$ & $0.002208$ & $0.008556$ & $0.000652$ \\
% $\Lambda_2$ & $0.000559$ & $0.002186$ & $0.005808$ & $0.001450$ \\
% $d_L$ & $0.001544$ & $\bm{0.01847}$ & $0.001273$ & $0.009157$ \\
% $\phi_c$ & $0.003500$ & $0.010714$ & $0.003338$ & $0.007874$ \\
% $\cos \iota$ & $0.001615$ & $0.012851$ & $0.006400$ & $0.009502$ \\
% $\psi$ & $0.004048$ & $0.011036$ & $0.001516$ & $0.005820$ \\
% $\alpha$ & $\bm{0.014008}$ & $0.001258$ & $\bm{0.009822}$ & $\bm{0.014136}$ \\
% $\sin \delta$ & $0.009570$ & $0.001761$ & $0.008934$ & $0.006362$ \\
% \hline\hline
%     \end{tabular*}
%     \caption{Jensen-Shannon divergences (in bits) between the marginal posterior obtained for GW170817 and GW190425 using \texttt{TaylorF2} and \texttt{IMRPhenomD\_NRTidalv2} with \textsc{Jim} and \textsc{pBilby}, with the highest value of each comparison in bold. The \ac{JS} divergences are bound between $[0, 1]$.}
%     \label{tab: JS divergences events}
% \end{table}

\subsection{Wall time}
\label{sec:wall time}

We report the total wall time spent on the \ac{PE} for the real and simulated events mentioned in this work in Tab.~\ref{tab: runtimes table}. We run \textsc{Jim} on a single NVIDIA A100-$40$ GB \ac{GPU}.

Each wall time of \textsc{Jim} mentioned below includes the time spent on the JIT compilation of the code and calculating the summary data used in relative binning, which on average takes around $2.5$ minutes.

For the analysis of the real events, the wall time of \textsc{Jim} additionally includes the time spent on calculating the reference parameters with \textsc{evosax} to initialize the relative binning likelihood, which takes between $2.5$ and $6$ minutes. Most of the wall time is spent on the training of the \ac{NF} after which the final production samples are produced in less than a minute. 

As a comparison, we also provide the runtimes of the injection-recovery tests of Sec.~\ref{sec:validation} in Tab.~\ref{tab: runtimes table}. For each set of injections, we report the median value of the runtimes. Since we set the reference parameters to the injected parameters, the runtime no longer includes the time spent on the \textsc{evosax} algorithm. We note that the median wall time for the simulated events analyzed with \texttt{TaylorF2} is slightly higher than that of the real events. This is because we have adjusted the sampling settings to be more robust to accommodate the broadly distributed simulated events in terms of signal-to-noise ratio and chirp masses. While the evaluation of the \texttt{IMRPhenomD\_NRTidalv2} waveform is slower than the \texttt{TaylorF2} waveform, we note that the speed of \textsc{Jim} is mainly determined by the convergence rate of the \ac{NF}, the value for the stopping criterion chosen by the user and the number of bins used in relative binning.  

We note that Ref.~\cite{Wong:2023lgb} achieved wall times of a few minutes for analyzing \ac{BBH} mergers with \textsc{Jim}. In our case, the wall time is higher, partly due to the increased dimensionality of the problem by including the tidal deformabilities, but mainly because of different sampler settings. In particular, our settings aim for robustness rather than speed by training the \ac{NF} until it achieves an average acceptance rate of $10\%$ ($20\%$) for the real events (injections). On the other hand, the settings chosen in Ref.~\cite{Wong:2023lgb} resulted in an average acceptance rate of $2\%$ -- $5\%$ for the \ac{NF}.

For the real events, the wall times are compared against their equivalent \textsc{pBilby}, \textsc{Bilby} with relative binning (RB-\textsc{Bilby}~\cite{Krishna:2023bug}), and \textsc{Bilby} with \ac{ROQ} (ROQ-\textsc{Bilby}~\cite{Morisaki:2023kuq})\footnote{Due to the unavailability of the \ac{ROQ} bases of \texttt{TaylorF2} with tidal effects and \texttt{IMRPhenomD\_NRTidalv2}, the bases of \texttt{IMRPhenomPv2\_NRTidalv2} are used.}, in Tab.~\ref{tab: runtimes table}.  \ac{ROQ} is a technique that creates an efficient representation of \ac{GW} approximants with a reduced set of basis elements that still accurately reconstruct the entire model space, thereby reducing the number of terms to be computed. These runs make use of nested sampling~\cite{Skilling:2006gxv}, in particular using \textsc{dynesty}~\cite{Speagle:2019ivv}, employing $1024$, $1000$ and $1000$ live points for \textsc{pBilby}, \textsc{RB-Bilby} and \textsc{ROQ-Bilby}, respectively. For the \textsc{RB-Bilby} runs, we use the same number of bins as the corresponding \textsc{Jim} runs, i.e., $200$ ($1000$) bins for the \texttt{TaylorF2} (\texttt{IMRPhenomD\_NRTidalv2}) waveform. This corresponds to setting the tunable $\epsilon$ parameter, as defined in Eq.~(10) of Ref.~\cite{Zackay:2018qdy}, to $\epsilon = 0.15$ ($\epsilon = 0.03$). The reference parameters are set to those found by the \textsc{evosax} algorithm from the corresponding \textsc{Jim} runs and are therefore not computed independently by \textsc{bilby}.

The \textsc{pBilby} runs are intended to verify the accuracy of our pipeline. However, \textsc{pBilby} is known to be expensive; therefore, we use $10$ Intel Skylake Xeon Platinum $8174$ \acp{CPU} ($480$ cores in total) per run so that the \ac{PE} completes in a reasonable amount of time. For RB-\textsc{Bilby} and ROQ-\textsc{Bilby}, we instead use a single Intel Xeon Silver $4310$ Processor \ac{CPU}, such that $24$ cores are used per run.\footnote{There are $12$ physical cores on an Intel Xeon Silver $4310$ Processor \ac{CPU}. On the machines used, there are two such \acp{CPU} installed. When running with 24 cores requested, the \acp{CPU} dynamically switches between with and without hyper-threading~\cite{intel_cpu_datasheet_4310}. For a conservative estimate of the energy consumption, we take only the \ac{TDP} of one \ac{CPU} in Sec.~\ref{sec: environmental impact}.}

\begin{table*}
    \centering
    \renewcommand{\arraystretch}{1.5}
    \begin{tabular*}{0.95\textwidth}{@{\extracolsep{\fill}} l l c c c c}
 Event & Waveform & \textsc{Jim} & \textsc{pBilby} & RB-\textsc{Bilby} & ROQ-\textsc{Bilby}  \\
 & & \footnotesize{($1$ GPU)} & \footnotesize{($480$ cores)} & \footnotesize{($24$ cores)} & \footnotesize{($24$ cores)} \\
  \hline\hline
 \multirow{2}{*}{GW170817} & \texttt{TF2} & $(4.85 + 15.33)$ min & $\phantom{0}9.64$ h & $3.8$ h & -- \\
 & \texttt{NRTv2} & $(5.38 + 25.59)$ min & $10.99$ h & $4.11$ h & $1.65$ h \\ \hline
\multirow{2}{*}{GW190425}  & \texttt{TF2} & $(2.63 + 18.30)$ min & $\phantom{0}8.18$ h & $2.81$ h & -- \\ 
 & \texttt{NRTv2} & $(3.26 + 21.20)$ min & $\phantom{0}4.91$ h & $2.42$ h & $0.97$ h \\ \hline
\multirow{2}{*}{Injection} & \texttt{TF2} & $\phantom{(0.000 + } 24.76\phantom{)}$ min & -- & -- & -- \\
& \texttt{NRTv2} & $\phantom{(0.000 + } 18.02\phantom{)}$ min & -- & -- & -- \\
\hline\hline
\end{tabular*}
\caption{Total wall time spent on conducting \ac{PE} on the events mentioned in this work, with the \texttt{TaylorF2} (\texttt{TF2}) and \texttt{IMRPhenomD\_NRTidalv2} (\texttt{NRTv2}) waveform models and using the resources mentioned in the main text for benchmarking. For the real events analyzed with \textsc{Jim}, we quote the time spent on \textsc{evosax} and \textsc{flowMC} separately. For the injections, we quote the median wall time. These wall times depend highly on the hardware used for conducting the analysis, i.e., one can achieve a shorter (longer) wall time if more (fewer) \acp{CPU} or \acp{GPU} are used.}
\label{tab: runtimes table}
\end{table*}

\subsection{Sampler efficiency}

Finally, we compare the efficiency of the different samplers by investigating the \ac{ESS} produced by the different pipelines averaged over the GW170817 and GW190425 runs. Since \textsc{bilby} makes use of nested sampling, the nested samples are rejection sampled to obtain posterior samples. Therefore, we take the total number of final posterior samples as \ac{ESS} values for the \textsc{bilby} runs. On the other hand, \textsc{Jim} relies on \ac{MCMC} sampling such that it directly produces posterior samples, which can however be correlated. For the \textsc{Jim} posteriors, we therefore compute the \ac{ESS} using Geyer's initial monotone sequence criterion~\cite{geyer_SSE_ref1, geyer_SSE_ref2} as implemented in the \textsc{Arviz} package~\cite{arviz_2019}. We estimate the \ac{ESS} for each parameter dimension and report the average across all parameters sampled during inference. The \textsc{Jim} posterior samples used for this calculation are taken only from \textsc{flowMC}'s production loop (i.e., after the \ac{NF} has been trained and its weights are frozen) and we discard the training samples for this analysis. Since the production loop is computationally efficient, this phase took less than $1$ minute for the \ac{GW} events considered here and can therefore be easily extended to produce more samples. The results are reported in Tab.~\ref{tab:number_of_samples}. We note that \textsc{Jim} can achieve a larger \ac{ESS} than other pipelines which can be attributed to the use of the global sampler. Indeed, since the \ac{NF} proposes samples independent of the history of each Markov chain, each accepted sample coming from the \ac{NF} is an effective sample. Therefore, the \ac{ESS} depends strongly on the global acceptance rate achieved in the \ac{GW} events considered here.

\begin{table}
    \centering
    \renewcommand{\arraystretch}{1.35}
    \begin{tabular*}{0.9\linewidth}{@{\extracolsep{\fill}} r c}
& Effective sample size \\
 \hline\hline
 \textsc{Jim} & $4.1 \times 10^4$ \\ \hline 
 \textsc{pBilby} & $6.5 \times 10^3$ \\ 
 \textsc{RB-Bilby} & $5.8 \times 10^3$ \\ 
 \textsc{ROQ-Bilby} & $7.4 \times 10^3$ \\ \hline\hline
    \end{tabular*}
    \caption{Comparison of the effective sample size (ESS), averaged over the \ac{GW} events considered in this work, between the different pipelines.}
    \label{tab:number_of_samples}
\end{table}

\section{Discussion}\label{sec:discussion}

Our numerical experiments from the previous sections demonstrate that \textsc{Jim} can robustly analyze \ac{BNS} mergers with high speed and without loss of accuracy. We discuss the implications of our pipeline for future work and compare our pipeline to other state-of-the-art pipelines that speed up \ac{PE}.

\subsection{Science cases}

The robustness and speed-up offered by \textsc{Jim} are crucial to meet the computational requirements of challenging science studies. For instance, while sky maps can be produced in low latency with \textsc{bayestar}~\cite{Singer:2015ema}, performing \ac{PE} on the intrinsic parameters improves the knowledge of the source properties and would be essential to improve estimates of electromagnetic (EM) emission, such as the brightness of a potential kilonova; see, e.g., Ref.~\cite{Stachie:2021noh}. In general, good guidance for EM follow-up searches is critical for improving the scientific return of follow-up campaigns while on the other hand lowering the requirement for expensive telescope time. As such, the speed of \textsc{Jim} in performing \ac{PE} offers a way to enhance low-latency follow-up strategies. 

Furthermore, the next generation of \ac{GW} detectors, such as Einstein Telescope~\cite{Punturo:2010zza} and Cosmic Explorer~\cite{Evans:2021gyd}, will have an overall greater sensitivity and broader frequency bandwidth compared to existing detectors \cite{Maggiore:2019uih, Evans:2023euw, Branchesi:2023mws}. Future detectors will observe more \ac{BNS} inspirals, which will also be in the sensitive band for a longer time, calling for efficient and effective \ac{PE} pipelines
\cite{Samajdar:2021egv,Pizzati:2021apa}. 

\subsection{Comparison to related works}\label{sec: comparison to related works}
Related works accelerating \ac{PE} can be mainly divided into likelihood-based methods, which directly evaluate the likelihood, and likelihood-free methods, which bypass the likelihood evaluation and instead rely on surrogates of the likelihood or posterior. We note that \textsc{Jim} belongs to the former class of methods.

One likelihood-based method to speed up \ac{PE} is relative binning and was discussed in Sec.~\ref{sec: relative binning}. Previous works have used relative binning for multi-messenger and low-latency \ac{PE} studies~\cite{Finstad:2020sok, Raaijmakers:2021slr}. Relative binning was also recently integrated into \textsc{Bilby}~\cite{Krishna:2023bug}.
 
Another way to speed up the likelihood evaluation is through \ac{ROM} and \ac{ROQ} methods. Using \ac{ROQ}, previous works have been able to perform \ac{PE} in the order of minutes~\cite{Canizares:2014fya, Morisaki:2020oqk, Morisaki:2023kuq}. However, \ac{ROQ} requires precomputing these reduced bases, which can be computationally expensive, as we show with our estimate in Appendix~\ref{appendix: ROQ estimate}. While these bases have to be constructed only once, \textsc{Jim}, on the other hand, does not require any precomputed quantities as input for the \ac{PE}.

Besides relative binning and \ac{ROQ}, other approximations of the likelihood have been studied as well, such as approximating the likelihood with Gaussian processes~\cite{Pankow:2015cra, Lange:2018pyp, Wysocki:2019grj} and a mesh-free interpolation method~\cite{Pathak:2022iar, Pathak:2023ixb}. Alternatively, one can marginalize over certain parameters in the likelihood to make parameter estimation more efficient and robust, such as in Ref.~\cite{Roulet:2024hwz}.

Apart from speeding up the evaluation of the likelihood, \ac{ML} can be used to improve the efficiency of the sampler and speed up \ac{PE}. For example, \textsc{nessai} accelerates nested sampling through normalizing flows~\cite{Williams:2021qyt}. Recently, this method achieved wall times similar to ours by combining \textsc{nessai} with the \ac{ROQ} approximation~\cite{Williams:2023ppp}. While similar in spirit to our work, we note that \textsc{Jim} makes use of \ac{MCMC} rather than nested sampling for exploring the likelihood landscape, and uses relative binning rather than \ac{ROQ} to speed up likelihood evaluation, thereby removing the need to precompute the \ac{ROQ} bases.

Finally, likelihood-free methods to accelerate \ac{PE} mainly consist of \ac{ML} models pretrained on simulated data to approximate the likelihood or posterior distributions~\cite{Gabbard:2019rde, Kolmus:2021buf, Kolmus:2024scm, Chua:2019wwt, Green:2020hst, Green:2020dnx, Dax:2021tsq, Dax:2022pxd, Bhardwaj:2023xph}. Contrary to these pipelines, \textsc{Jim} conducts the \ac{PE} without the need for pretraining.\\

\subsection{Environmental impact}\label{sec: environmental impact}

Besides a reduction in wall time, our setup additionally offers an ecologically desirable implementation to perform \ac{PE}. 

We compare the carbon footprint of \textsc{Jim} relative to the other pipelines reported in Tab.~\ref{tab: environmental impact}. The \ac{TDP} of an NVIDIA A100-$40$ GB GPU, used for the \textsc{Jim} runs, is $400$ W~\cite{nvidia_a100_datasheet}. The \textsc{pBilby} runs used a single Intel Xeon Platinum 8174 Processor with a \ac{TDP} of $240$ W~\cite{ark_intel_cpu_datasheet}. The RB-\textsc{Bilby} and ROQ-\textsc{Bilby} used a single Intel Xeon Silver 4310 Processor CPU with a \ac{TDP} of $120$ W~\cite{intel_cpu_datasheet_4310}. 

Based on our \ac{PE} runs, we take the average wall time for each pipeline, and the \ac{TDP} reported above to estimate the required energy to produce the results for all $200$ injection runs and the $4$ real event runs shown in this work. The results are reported in amounts of kWh in Tab.~\ref{tab: environmental impact}.\footnote{The obtained numbers have been rounded for clarity and ease of presentation.} In order to make a fair comparison against \textsc{Jim}, we also report the time taken up by the pretraining phase of the \ac{ROQ} method, i.e., the time taken to construct the \ac{ROQ} bases. The details of this estimate can be found in Appendix~\ref{appendix: ROQ estimate}. For comparison, the average annual electricity consumption of a Netherlands household in 2021 was ${2810}$ kWh~\cite{Statistics_Netherlands_2023}.

To translate these into tangible numbers, we will assume that $0.328$ kg $\rm{CO}_2$ is produced per kWh~\cite{NL_CO2_emissiefactoren} and that it approximately takes $50$ trees a year to capture ${1000}$ kg of $\rm{CO}_2$ \cite{climateneutral_trees_reference}. Given a year, it takes around $0.55 \rm{\ trees}$ to absorb the amount of $\rm{CO}_2$ generated by \textsc{Jim}. On the other hand, around $67.68$ and $1.32\rm{\ trees}$ are needed for absorbing the $\rm{CO}_2$ generated by \textsc{pBilby} and RB-\textsc{Bilby}, respectively. While only $0.52 \rm{\ trees}$ are needed to counter the carbon footprint from the sampling of \ac{ROQ}-\textsc{Bilby}, the precompute of the \ac{ROQ} bases requires around $0.44\rm{\ trees}$. We can, therefore, conclude that \textsc{Jim} is more environmentally friendly than the other pipelines considered in performing the \ac{PE} runs in this work. However, we would like to emphasize that the cost associated with building the \ac{ROQ} bases is a one-time cost only. Therefore, we estimate that \ac{ROQ}-\textsc{Bilby} will have a similar ecological footprint as \textsc{Jim} after around $3000$ \ac{PE} runs.

\begin{table}
    \centering
    \renewcommand{\arraystretch}{1.35}
    \begin{tabular*}{0.975\linewidth}{@{\extracolsep{\fill}} l c r r r}
& & kWh & $\rm{CO}_2$ [$10^3$ kg] & Trees${}^\dagger$ \\
 \hline\hline
 \textsc{Jim} & & $\phantom{00}34$ & $\phantom{0}11$ & $\phantom{000}0.55$ \\ \hline 
 \textsc{pBilby} & & $4127$ & $1354$ & $67.68$ \\ \hline 
 \textsc{\textsc{RB-Bilby}} & & $80$ & $26$ & $1.32$ \\ \hline 
 \multirow{2}{*}{\textsc{ROQ-Bilby}} & sampling & $32$ & $10$ & $0.52$ \\ 
 & precompute${}^\ddagger$ & $27$ & $9$ & $0.44$ \\
\hline\hline
    \end{tabular*}
    \caption{Estimate of the environmental impact of performing all runs in this work with different pipelines. ${}^\dagger$Number of trees needed to capture the emitted $\rm{CO}_2$ in a year. ${}^\ddagger$Based on our estimate of the resources needed to build the required \ac{ROQ} bases, see Appendix~\ref{appendix: ROQ estimate}.}
    \label{tab: environmental impact}
\end{table}

\subsection{Future work}
Finally, we mention a few directions in which we wish to pursue the developmental work of our pipeline.

While our current approach, as outlined in Sec.~\ref{sec: relative binning}, allows us to use \texttt{IMRPhenomD\_NRTidalv2} with the existing relative binning implementation, it is desirable to extend the relative binning method to waveforms which include a tapering window, such as the \texttt{NRTidal} family of waveforms. For this, one requires a scheme to construct the bins without making prior assumptions regarding the waveforms used. An example of such an agnostic approach is given by Ref.~\cite{Narola:2023men}. 

In addition, a current bottleneck in the development of \textsc{Jim} is the need for a \textsc{jax}-compatible implementation of existing \ac{GW} approximants. At the time of writing, a \textsc{ripple} implementation of the precessing waveforms \texttt{IMRPhenomPv2} and \texttt{IMRPhenomXPHM} are under development. Recently, a \textsc{jax}-compatible implementation of the \texttt{IMRPhenomPv2\_NRTidal} approximant has been constructed~\cite{Dax:2024mcn}. Additionally, we will investigate the possibility of training surrogate \ac{ML} models to approximate the waveforms, which removes the need to re-implement waveforms from scratch and reduces development time.  

Furthermore, future work can investigate the integration with existing packages. While our current pipeline produces posterior samples of any desired target distribution, it does not output the evidence, which is crucial for model selection. In the future, we wish to estimate the Bayesian evidence by performing importance sampling on our produced samples. Future work can integrate \textsc{Jim} with \textsc{denmarf}~\cite{lo2023denmarf} or \textsc{harmonic}~\cite{mcewen2023machine, SpurioMancini:2022vcy, polanska2024learned} to obtain the Bayesian evidence. Secondly, one can attempt to integrate the recently introduced package \textsc{astreos}~\cite{McGinn:2024nkd} with our pipeline. \textsc{astreos} can generate confidence intervals for the equations of state in less than one second by relying on normalizing flows. However, the \ac{NF} maps the component masses and tidal deformabilities to a set of so-called auxiliary parameters from which the \ac{EOS} constraints can be inferred. Obtaining estimates for these intrinsic parameters can be done reliably and efficiently with \textsc{Jim}, which can therefore empower \textsc{astreos}. 

\section{Conclusions}\label{sec:conclusions}

In this work, we have demonstrated the robustness, accuracy, and speed of \textsc{Jim} in performing parameter estimation for gravitational waves of binary neutron star mergers with tidal effects. By combining relative binning, \textsc{jax}, gradient-based \ac{MCMC} and normalizing flows, \textsc{Jim} can produce posteriors accurately in the order of minutes without requiring any precomputed input such as a pretrained normalizing flow. Specifically, we are able to analyze GW170817 and GW190425 with the \texttt{TaylorF2} and \texttt{IMRPhenomD\_NRTidalv2} waveforms in $20$ to $31$ minutes of total wall time, depending on the sampling settings and the number of bins used in relative binning. Evaluated on a large suite of simulated events, the median total wall times are below $30$ minutes for both waveforms.  As such, \textsc{Jim} can be an indispensable tool for future science cases, such as low-latency follow-up strategies and data analysis with next-generation gravitational wave detectors. Additionally, we have shown that \textsc{Jim} has significantly reduced the environmental impact of parameter estimation as compared to other pipelines. Therefore, \textsc{Jim} presents an ecologically sustainable approach to addressing these computationally intensive tasks.

\appendix 
\FloatBarrier

\section{Data availability}

All code used to produce the figures in this paper is available at the following link: \url{https://github.com/ThibeauWouters/TurboPE-BNS}. Posterior samples will be shared upon request.

\section{Corner plots}
\label{sec:corner_plot}
\begin{figure*}
    \centering
    \includegraphics[width=\textwidth]{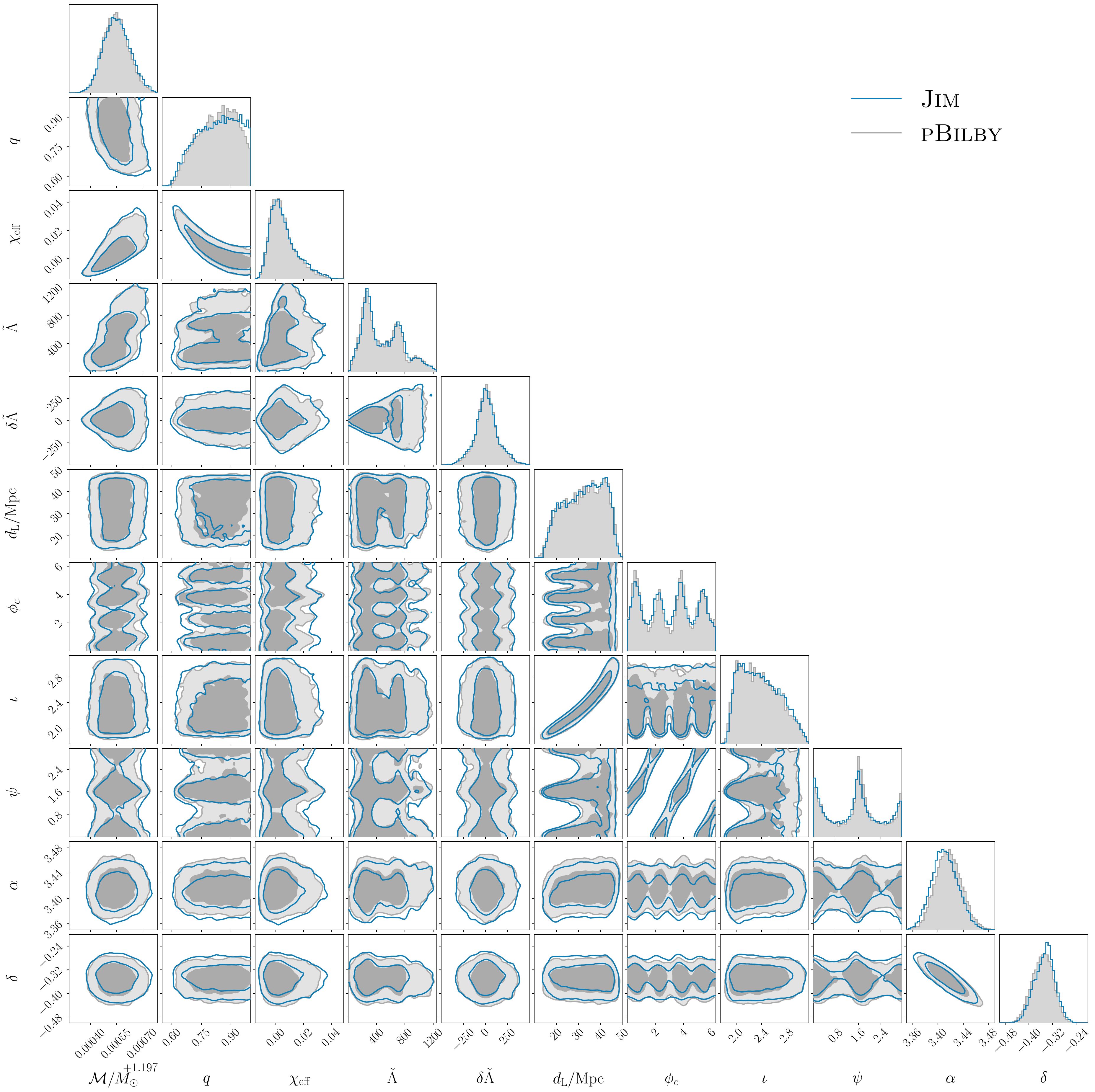}
    \caption{Comparison of the posterior distributions of GW170817, obtained with the \texttt{TaylorF2} waveform, using \textsc{Jim} and \textsc{pBilby}.}
    \label{fig:GW170817_TaylorF2}
\end{figure*}

\begin{figure*}
    \centering
    \includegraphics[width=\textwidth]{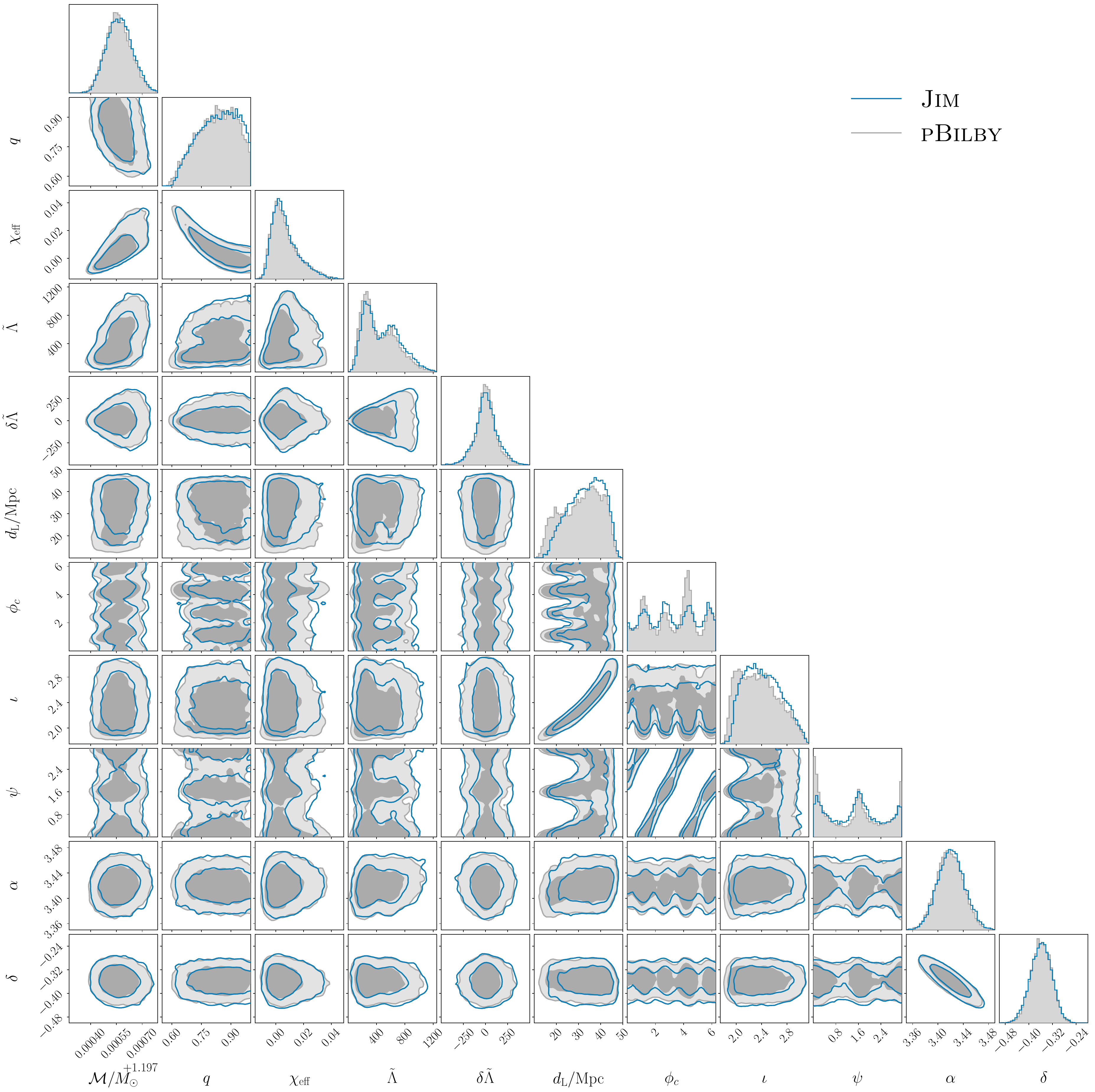}
    \caption{Comparison of the posterior distributions of GW170817, obtained with the \texttt{IMRPhenomD\_NRTidalv2} waveform, using \textsc{Jim} and \textsc{pBilby}.}
    \label{fig:GW170817_NRTidalv2}
\end{figure*}

\begin{figure*}
    \centering
    \includegraphics[width=\textwidth]{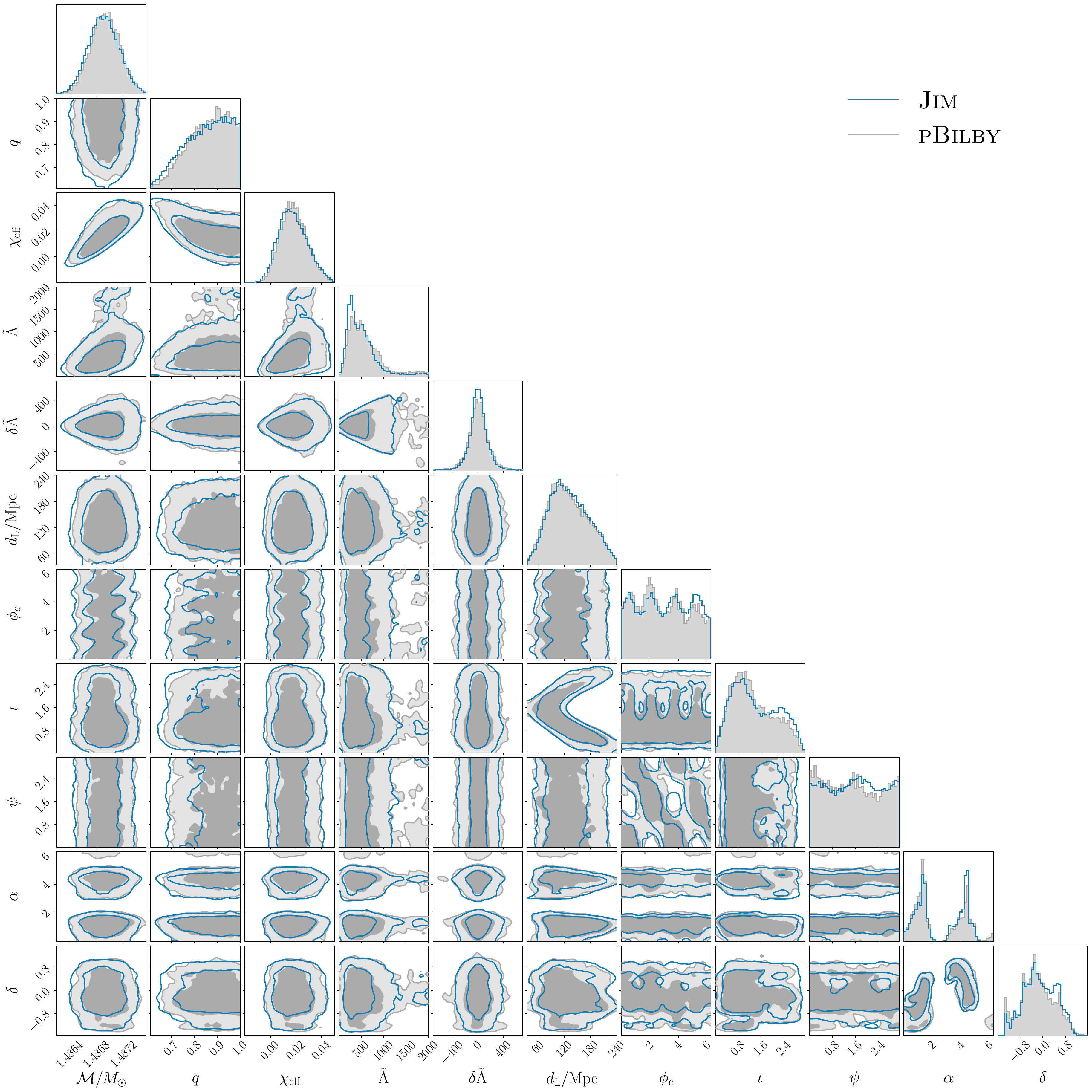}
    \caption{Comparison of the posterior distributions of GW190425, obtained with the \texttt{TaylorF2} waveform, using \textsc{Jim} and \textsc{pBilby}.}
    \label{fig:GW190425_TaylorF2}
\end{figure*}

\begin{figure*}
    \centering
    \includegraphics[width=\textwidth]{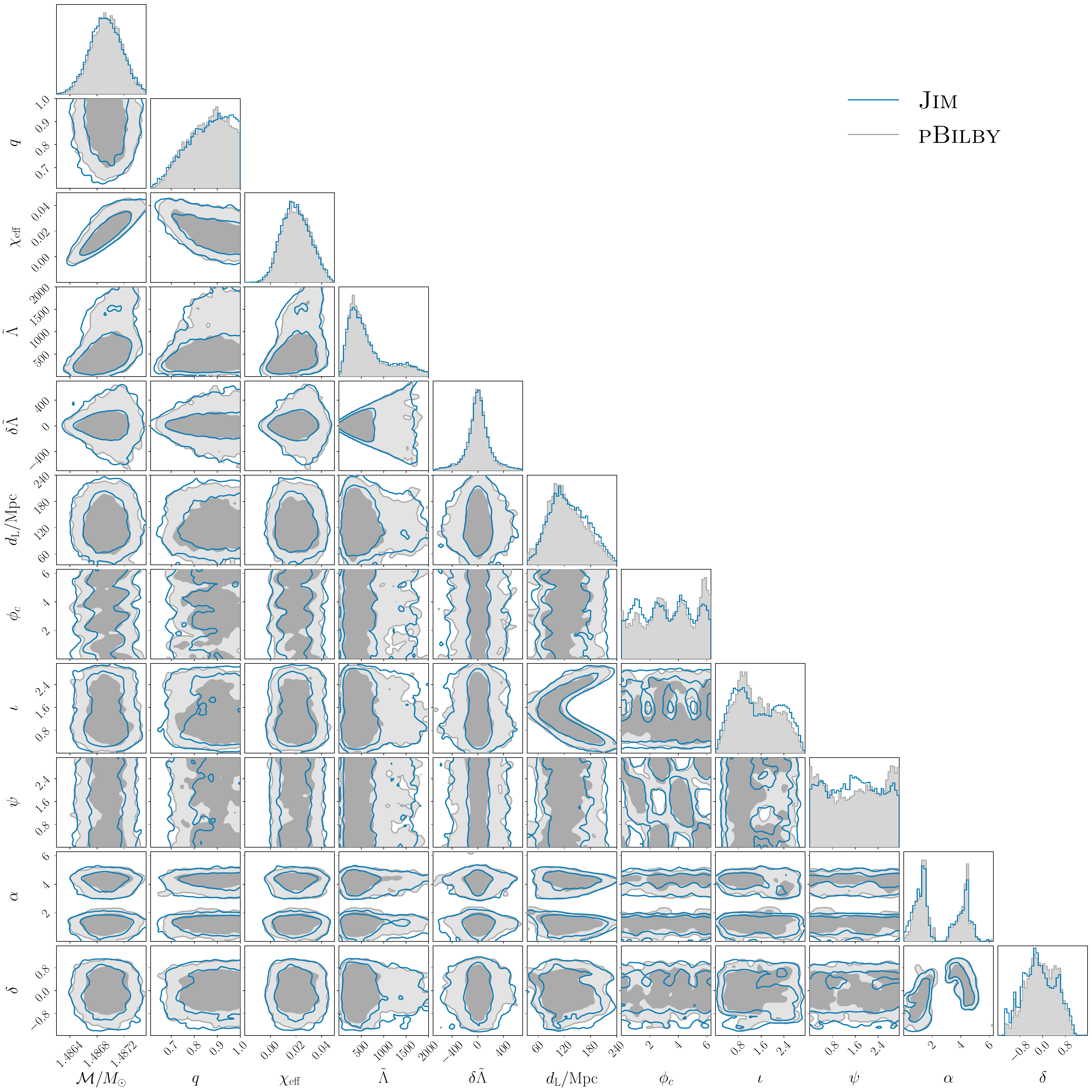}
    \caption{Comparison of the posterior distributions of GW190425, obtained with the \texttt{IMRPhenomD\_NRTidalv2} waveform, using \textsc{Jim} and \textsc{pBilby}.}
    \label{fig:GW190425_NRTidalv2}
\end{figure*}

We show the corner plots of the posterior distributions obtained from \textsc{Jim} and \textsc{pBilby} in~\Cref{fig:GW170817_TaylorF2,fig:GW170817_NRTidalv2,fig:GW190425_TaylorF2,fig:GW190425_NRTidalv2} for the GW170817 and GW190425 events analyzed with \texttt{TaylorF2} and \texttt{IMRPhenomD\_NRTidalv2}. Instead of the individual aligned spins, we plot the $\chi_{\rm{eff}}$ parameter as defined by Eq.~\eqref{eq:effective spin}. 
Moreover, we show the tidal deformabilities in terms of $\tilde{\Lambda}$ and $\delta\tilde{\Lambda}$, defined by
\begin{equation}
\begin{aligned}
\tilde{\Lambda} &= \frac{16}{13}\frac{(m_1 + 12 m_2)m_1^4\Lambda_1 + (m_2 + 12 m_1)m_1^4\Lambda_2}{(m_1+m_2)^5},\\
\delta\tilde{\Lambda} &= \left(\frac{1690}{1319}\eta - \frac{4843}{1319}\right)\frac{m_1^4\Lambda_1 - m_2^4\Lambda_2}{{(m_1+m_2)^4}}\\
&+\frac{6162}{1319}\sqrt{1 - 4\eta}\frac{m_1^4\Lambda_1 + m_2^4\Lambda_2}{{(m_1+m_2)^4}},
\end{aligned}
\end{equation}
where $\eta=m_1m_2/(m_1 + m_2)^2$ is the symmetric mass ratio. The plotted contours show the $1\sigma$ and $2\sigma$ significance levels. Across the four sets of posterior samples shown, we observe qualitative agreement between the ones obtained from \textsc{Jim} and those obtained from \textsc{pBilby}.

\FloatBarrier

\section{Estimate of \ac{ROQ} environmental impact}\label{appendix: ROQ estimate}

We provide details on how the estimate of the environmental impact of constructing \ac{ROQ} bases, reported in Tab.~\ref{tab: environmental impact}, is computed based on Ref.~\cite{Soichiro_mail}.

Building the \ac{ROQ} bases for the \texttt{IMRPhenomPv2\_NRTidalv2} waveform takes around $\sim1$ day when using $\sim100$ cores on the NEMO cluster owned by UWM when making use of multi-banding~\cite{Morisaki:2021ngj} to speed up the building of the \ac{ROQ} bases. We will assume around $\sim5$ Intel Xeon Gold 6136 Processor \acp{CPU} were used, each having a \ac{TDP} of $150$ W~\cite{intel_cpu_datasheet_gold_8380}. Therefore, we estimate the energy consumption of building the \ac{ROQ} bases for this waveform to be around $27$ kWh.

Currently, there is no \ac{ROQ} implementation of the \texttt{TaylorF2} waveform with tidal effects publicly available. To estimate the computational cost to build the \ac{ROQ} bases for this waveform, we will conservatively assume that this requires around half of the resources used for the \texttt{IMRPhenomPv2\_NRTidalv2} as the computational cost depends on the complexity of the waveform model. 

We would like to emphasize that this is only an estimate of the computational cost. In particular, further advancements in the techniques used for building \ac{ROQ} bases can potentially reduce the computational cost, e.g. Ref.~\cite{Morras:2023pug}. 

\section*{Acknowledgments}
We thank Justin Janquart, Harsh Narola, Anna Puecher, Kaze Wong, Thomas Edwards, Nihar Gupte,  Michael Williams, Colm Talbot, Lalit Pathak and Soichiro Morisaki for fruitful discussions and feedback that led to the improvement of this work.
T.W., P.T.H.P., and C.V.D.B. are supported by the research program of the Netherlands Organization for Scientific Research (NWO).
T.W., P.T.H.P., T.D.\ acknowledge support from the Daimler and Benz Foundation for the project ``NUMANJI''. T.D.\ acknowledges support from the European Union (ERC, SMArt, 101076369). Views and opinions expressed are those of the authors only and do not necessarily reflect those of the European Union or the European Research Council. Neither the European Union nor the granting authority can be held responsible for them.
We thank SURF (www.surf.nl) for the support in using the National Supercomputer Snellius under project number EINF-6587 and EINF-8596.
Computations have been performed on the SuperMUC-NG (LRZ) under project number pn56zo.
The authors acknowledge the computational resources provided by the LIGO Laboratory's CIT cluster, which is supported by National Science Foundation Grants PHY-0757058 and PHY0823459.
This research has made use of data or software obtained from the Gravitational Wave Open Science Center (gwosc.org), a service of the LIGO Scientific Collaboration, the Virgo Collaboration, and KAGRA. This material is based upon work supported by NSF's LIGO Laboratory which is a major facility fully funded by the National Science Foundation, as well as the Science and Technology Facilities Council (STFC) of the United Kingdom, the Max-Planck-Society (MPS), and the State of Niedersachsen/Germany for support of the construction of Advanced LIGO and construction and operation of the GEO600 detector. Additional support for Advanced LIGO was provided by the Australian Research Council. Virgo is funded, through the European Gravitational Observatory (EGO), by the French Centre National de Recherche Scientifique (CNRS), the Italian Istituto Nazionale di Fisica Nucleare (INFN) and the Dutch Nikhef, with contributions by institutions from Belgium, Germany, Greece, Hungary, Ireland, Japan, Monaco, Poland, Portugal, Spain. KAGRA is supported by Ministry of Education, Culture, Sports, Science and Technology (MEXT), Japan Society for the Promotion of Science (JSPS) in Japan; National Research Foundation (NRF) and Ministry of Science and ICT (MSIT) in Korea; Academia Sinica (AS) and National Science and Technology Council (NSTC) in Taiwan.

\appendix

\begin{acronym}
    \acro{PE}[PE]{parameter estimation}
    \acro{GW}[GW]{gravitational wave}
    \acrodefplural{GWs}{gravitational waves}
    \acro{CBC}[CBC]{compact binary coalescences}
    \acro{NS}[NS]{neutron star}
    \acrodefplural{NSs}{neutron stars}
    \acro{MCMC}[MCMC]{Markov chain Monte Carlo}
    \acro{HMC}[HMC]{Hamiltonian Monte Carlo}
    \acro{MALA}[MALA]{Metropolis-adjusted Langevin algorithm}
    \acro{NF}[NF]{normalizing flow}
    \acro{BBH}[BBH]{binary black hole}
    \acro{BNS}[BNS]{binary neutron star}
    \acro{NSBH}[NSBH]{neutron star-black hole}
    \acro{EOS}[EOS]{equation of state}
    \acro{PN}[PN]{post-Newtonian}
    \acro{EOB}[EOB]{effective one-body}
    \acro{IMRPhenom}[\texttt{IMRPhenom}]{inspiral-merger-ringdown phenomenological}
    \acro{JS}[JS]{Jensen-Shannon}
    \acro{p-p}[p-p]{percentile-percentile}
    \acro{CPU}[CPU]{central processing unit}
    \acro{GPU}[GPU]{graphical processing unit}
    \acro{TPU}[TPU]{tensor processing unit}
    \acro{TDP}[TDP]{thermal design power}
    \acro{kWh}[kWh]{kilowatt-hour}
    \acro{ML}[ML]{machine learning}
    \acro{ROQ}[ROQ]{reduced order quadrature}
    \acro{ROM}[ROM]{reduced order modelling}
    \acro{SNR}[SNR]{signal-to-noise ratio}
    \acro{LF}[LF]{low-frequency}
    \acro{HF}[HF]{high-frequency}
    \acro{ASD}[ASD]{amplitude spectral density}
    \acro{PSD}[PSD]{power spectral density}
    \acro{FIM}[FIM]{Fisher information matrix}
    \acro{SSM}[SSM]{subsolar mass}
    \acro{ESS}[ESS]{effective sample size}
    \acro{RNG}[RNG]{random number generator}
\end{acronym}

\bibliography{references}{}
\bibliographystyle{apsrev4-1}

\end{document}